\documentclass[12pt]{iopart}
\usepackage[xdvi]{graphicx}%

\newcommand{\bv}[1]{{\bf #1}}

\begin{document}

\title{Long-time tails for sheared fluids}
\author{Michio Otsuki$^1$ and Hisao Hayakawa$^2$}

\address {
$^1$ Department of Physics and Mathematics, Aoyama Gakuin University,
5-10-1 Fuchinobe, Sagamihara, Kanagawa, 229-8558, Japan. }
\address{  $^2$ Yukawa Institute for Theoretical Physics, Kyoto University,  Kitashirakawaoiwake-cho, Sakyo-ku, Kyoto 606-8502, Japan.}


\ead{\mailto{otsuki@phys.aoyama.ac.jp}}

\begin{abstract}
The long-time behaviors of the velocity autocorrelation function (VACF)
for sheared fluids  
are investigated theoretically and numerically. 
It is found the existence of the cross-overs of
VACF from $t^{-d/2}$ to $t^{-d}$  
in sheared fluids of elastic particles without any thermostat,  and
from $t^{-d/2}$ to $t^{-(d+2)/2}$ 
in both sheared fluids of elastic particles with a thermostat and sheared granular fluids,
where $d$ is the spatial dimension. 
The validity of the predictions has been confirmed by
our numerical simulations.
\end{abstract}

\maketitle

The slow relaxation of the current autocorrelation functions
to the equilibrium state is one of the most important characteristics
in nonequilirbium statistical physics \cite{alder}--\cite{ernst05}.
It is known that the long-time tails are caused by 
the correlated collisions 
which cause anomalous transport behaviors such as the system size dependence of 
the transport coefficients \cite{Narayan,Nishino}.
The analysis of such behaviors is still an active subject in nonequilibrium physics. 
Although there is the long-time tail  $t^{-d/2}$ with the time $t$ and the spatial dimension $d$ 
even at equilibrium, the tail under an external force such as a steady shear has
different feature from that at equilibrium.
Indeed, we believe that the viscosity $\eta$ 
depends on the absolute value of the shear rate
$\dot\gamma$ as $\eta-\eta_0=\eta'\log(\dot\gamma)$ for $d=2$ and $\eta-\eta_0=\eta''\dot\gamma^{1/2}$ for $d=3$ in the sheared ordinary
 fluids at a constant temperature,
where $\eta_0$ is the Newtonian viscosity at equilibrium, and $\eta'$ and $\eta''$ are proportional 
constants \cite{Yamada,Lutsko}.
These results are obtained from the assumption that
the current autocorrelation functions for $t>\dot{\gamma}^{-1}$ decay much
faster than the conventional tail. 
Nevertheless, because of the lack of systematic studies in the long time region, we arise the following question: 
(i) {\it What are the long-time behaviors of the autocorrelation functions 
for $t>\dot{\gamma}^{-1}$}? 

The system becomes unsteady because of the increment of the temperature 
due to the viscous heating effect when we add the shear to a system consisting of elastic particles.
We, then, sometimes introduce a thermostat to keep an isothermal steady state of the fluid under the shear \cite{Evans}--\cite{Evans93}.
We, however, do not know the details of roles of thermostats
in the sheared fluids.
   On the other hand, the sheared granular fluid in which there are inelastic collisions between particles 
can be regarded as one type of isothermal fluids, because
the sheared granular fluid can keep a constant temperature
under the balance between the shear and the dissipation due to inelastic collisions.
Indeed, our recent paper has confirmed that an unified description 
of both sheared granular fluids and sheared isothermal fluids with a thermostat
can be used for equal-time long-range correlations as long as the systems keep uniform shear flows \cite{Otsuki09:LR}.
We, thus, arise the second question: (ii) 
{\it What are actual relations among the sheared fluid of elastic particles without any thermostats, the sheared isothermal fluid
of elastic particles and the sheared granular fluid}? 

The interest in the role of the long-tails in granular fluids is rapidly 
growing \cite{Kumaran}--\cite{Hayakawa09}.
In particular, Kumaran suggested the existence of 
an interesting fast decay of the 
autocorrelation functions, $t^{-3d/2}$ in sheared dense granular fluids 
\cite{Kumaran,Kumaran09}, 
while Otsuki and Hayakawa predicted that the correlation of the shear stress
  decays as $t^{-(d+2)/2}$ \cite{Otsuki09}.
These predictions are nontrivial, but the exponents for the tails 
are not confirmed in the recent experiments \cite{Orpe08}.
On the other hand, the existence of conventional long-time tails in velocity autocorrelation function  (VACF) has been confirmed, 
while the absence of long-tail in the correlation of the heat flux is found
 \cite{Hayakawa}.
Thus, we encounter the third question: 
(iii) {\it What is the true long-time tail in the sheared granular fluids}?

In this letter, to answer the above three questions we study the long-time behaviors of VACFs in sheared elastic particles with or without thermostats and sheared granular particles.
Our theoretical method is based on the classical one developed by Ernst et al. \cite{ernst71,Hayakawa}, and the theoretical
results will be verified from our numerical simulations.

Let us consider
the system consists of $N$ identical smooth and hard spherical particles with the mass
$m$ and 
the diameter $\sigma$ in the volume $V$. 
The position and the velocity of the $i$-th particle at time $t$
are $\bv{r}_i(t)$ and $\bv{v}_i(t)$, respectively.  
The particles collide instantaneously with each other
with a restitution constant $e$ which is equal to unity for elastic particles
and less than unity for granular particles.
When the particle $i$ with the velocity $\bv{v}_i$ collides with the particle  $j$ with $\bv{v}_j$,
the post-collisional velocities $\bv{v}^*_i$ and $\bv{v}^*_j$ are respectively given by
$\bv{v}^*_i  =  \bv{v}_i  - \frac{1}{2} (1+e) (\bv{\epsilon} \cdot
\bv{v}_{ij}) \bv{\epsilon} $ and
$\bv{v}^*_j  =  \bv{v}_j  + \frac{1}{2} (1+e) (\bv{\epsilon} \cdot
\bv{v}_{ij})\bv{\epsilon} $,
where $\bv{\epsilon}$ is the unit vector parallel to the relative position of the
two colliding particles at contact, and $\bv{v}_{ij}=\bv{v}_{i}-\bv{v}_{j}$. 

We assume that the velocity profile is given by
$c_{\alpha}(\bv{r}) = \dot{\gamma} y \delta_{\alpha,x}$ in our system,
where the Greek suffix $\alpha$ denotes the Cartesian component.
In this letter, we discuss the following situations: (a) A sheared system 
of elastic particles
without any thermostat, (b)  a sheared system of elastic particles with 
the velocity rescaling thermostat, and (c) 
a sheared granular system with the restitution constant $e<1$.
We abbreviate them the sheared heating (SH),  the sheared thermostat (ST),  
and the sheared granular (SG) systems for later discussion.
To avoid the consideration of the contribution from the potential terms,  we assume that
the sheared fluid consists of  a dilute gas of particles.

We are interested in VACFs for $d$-dimensional systems
\cite{Cummings, Campbell}
\begin{equation}
C^{(d)}_\alpha(t) \equiv \frac{1}{N}\sum_{i=1}^N \langle  
v_{ i, \alpha}'(0) v_{i, \alpha}'(t)
\rangle, \quad C^{(d)}(t) = \frac{1}{d} \sum_{\alpha=1}^d C_\alpha(t), \label{C:def}
\end{equation}
where we have introduced 
$v_{i, \alpha}'(t)\equiv v_{ i, \alpha}(0) - c_\alpha(\bv{r}_i(0) )$,
and $C^{(d)}(t)$ is the superposition of $C^{(d)}_\alpha(t)$.
VACFs are expected to be represented 
by the fluctuation of hydrodynamic fields  around uniform shear flow characterized by
the uniform shear velocity $c_{\alpha}(\bv{r})$ and
the homogeneous temperature $T_H$ \cite{ernst71,Hayakawa} as
\begin{equation}
C^{(d)}_\alpha(t) \simeq 
\int d\bv{v}'_0 f_0(v'_0)v'_{0,\alpha} \int \frac{d\bv{k}}{(2\pi)^d}u_{\bv{k},\alpha}(t)P_{-\bv{k}}(t) ,
 \label{C2:eq}
\end{equation}
where $u_{\bv{k},\alpha}(t)$, $P_{\bv{k}}(t)$, $\bv{v}'_0$ and $f_0(v'_0)$  are  
the Fourier transforms of 
$\alpha$-component of the velocity field and the 
probability distribution  function of a tracer particle,
the initial peculiar velocity of a particle defined by
$\bv{v}'_0 = \bv{v}_0 - \bv{c}(\bv{r}_0)$ with
the initial velocity $\bv{v}_0$, 
 and the initial velocity distribution function, respectively.
Here, we should note that the wave number $\bv{k}$ is taken
to satisfy the Lees-Edwards boundary condition \cite{onuki}.
The validity of eq. (\ref{C2:eq}) in 
 the nonequilibrium situations has been verified in Ref. \cite{Hayakawa}.
Thus, we need to estimate the behaviors of $\bv{u}_{\bv{k}}(t)$ and $P_{\bv{k}}(t)$ to obtain VACFs.

As was shown in the previous studies \cite{ernst71,Hayakawa}, we adopt the linearized hydrodynamics of fluctuating fields
 around the uniform shear flow to evaluate VACFs.
As in the equilibrium case, we decompose 
$\bv{u}_{\bv{k}}(t)$ into the longitudinal mode $u_{\bv{k}\parallel}(t)$ and 
 the transverse mode $u^{(i)}_{\bv{k}\perp}(t)$ 
 as $\bv{u}_{\bv{k}}(t) = u_{\bv{k}\parallel}(t)\bv{e}_\parallel+ 
 \sum_i^{d-1} u^{(i)}_{\bv{k}\perp}(t)\bv{e}^{(i)}_\perp$, 
 where we have introduced
 $\bv{e}_\parallel=\bv{k}(-\dot \gamma t)/k(-\dot \gamma t)$,
 and 
$\bv{e}^{(i)}_\perp= \{ \bv{e}_{\alpha_i} 
- (\bv{e}_\parallel \cdot \bv{e}_{\alpha_i}) \bv{e}_\parallel
- \sum_j^{i-1}(\bv{e}^{(j)}_\perp \cdot \bv{e}_{\alpha_i}) \bv{e}^{(j)}_\perp \}
/ {\cal N}$ with
$\bv{k}(\dot \gamma t)= \bv{k} + \dot{\gamma}t k_x \bv{e}_y$ \cite{onuki} and
 ${\cal N} = | \bv{e}_{\alpha_i} 
- (\bv{e}_\parallel \cdot \bv{e}_{\alpha_i}) \bv{e}_\parallel
- \sum_j^{i-1}(\bv{e}^{(j)}_\perp \cdot \bv{e}_{\alpha_i}) \bv{e}^{(j)}_\perp|$
.
Here, 
$\bv{e}_{\alpha_i}$ is the unit vector parallel to  $\alpha_i$
direction.
In this letter, we take $\alpha_1$ as $y$.
We introduce 
$w^{(i)}_{\bv{k}\perp}(t) \equiv u^{(i)}_{\bv{k}\perp}(t)/\sqrt{T_H(t)}$,
and 
the time evolution equation of $w^{(i)}_{\bv{k}\perp}(t)$ is given by
\begin{eqnarray}
\partial_t  w^{(i)}_{\bv{k}\perp}(t)&=&
-\nu^*\sqrt{T_H(t)} k(-\dot \gamma t) ^2 w^{(i)}_{\bv{k}\perp}(t)
+ \dot \gamma F(\bv{k}(-\dot \gamma t),t), 
\label{u_kp}
\end{eqnarray}
where  we have introduced
$F(\bv{k},t) = (\bv{e}^{(i)}_\perp \cdot \bv{e}_x) 
u_{\bv{k} y}(t)/\sqrt{T_H(t)} +
\delta_{i,1} k_x k_\perp / k
u_{\bv{k}\parallel }(t) /\sqrt{T_H(t)} 
+ \nu^*\{ ik_x(\bv{e}^{(i)}_\perp \cdot \bv{e}_y) +
 ik_y
 (\bv{e}^{(i)}_\perp \cdot \bv{e}_x) \} T_{\bv{k}}(t)/T_H(t)$ with 
 $k_\perp=\sqrt{k^2 - k_y^2}$.
Here, $T_{\bv{k}}(t)$ is the Fourier transform of the temperature,
and $\nu^*$ is unimportant part of the kinetic viscosity
which is independent of the temperature.

We note that the transverse mode  $w^{(i)}_{\bv{k}\perp}(t) $
is not independent of the longitudinal modes for sheared fluids
because of the existence of  $F(\bv{k},t)$. 
Nevertheless, 
 we can ignore the mixing effect
by assuming the small shear rate
  satisfying $\dot\gamma \ll \nu^*\sqrt{T_H(t)} k(-\dot \gamma t)^2$.
It should be noted that
the condition $\dot\gamma \ll \nu^*\sqrt{T_H(t)} k(-\dot \gamma t)^2$
is not satisfied in the limit $k \rightarrow 0$.
However,  $k$ has the infrared cutoff $2 \pi / L$ with the system size $L$
for the sheared system. 
Thus, the condition
 can be used  when we assume small enough $\dot \gamma$
 or high enough initial temperature $T_H(0)$ for SH and ST,
and $1-e^2\ll 1$ for SG,
as in Ref. \cite{Otsuki09:LR}.

We expect that 
the longitudinal mode proportional to $\bv{e}_\parallel$ 
is  irrelevant for sheared fluids 
 because of the existence of the sound wave in sheared fluids 
even when we consider SG. This is contrast to the case of
freely cooling granular fluids \cite{Hayakawa}.

The time evolution equation of 
$P_{\bv{k}}(t)$ in 
sheared fluids is
\begin{eqnarray}
\partial_t
P_{\bv{k}}(t)&=&-D^*\sqrt{T_H(t)} k(-\dot \gamma t)^2 P_{\bv{k}}(t),
\label{P_k}
\end{eqnarray}
where $D^*$ is 
 unimportant part of 
the diffusion coefficient, which is independent of the temperature.

From eqs. (\ref{u_kp}) and (\ref{P_k}), it is obvious that the time evolution of the homogeneous temperature $T_H(t)$ plays a key role in the long time
behavior of VACF. When we consider a system of SH, the temperature 
increases by the viscous heating. In this case, the temperature obeys
$dT_H/dt=2m \nu^*\sqrt{T_H}\dot{\gamma}^2/d$ and its solution is $\sqrt{T_H(t)}=\sqrt{T_H(0)}+m\nu^*\dot{\gamma}^2 t/d$.
On the other hand, $T_H(t)$ should be a constant for either ST or SG
from the balance between the viscous heating and the energy dissipation.
Therefore, we reach an important and an unexpected conjecture
that the behavior of VACFs for nearly elastic SG is the same as that for ST,
 but is different from that for SH.
The difference between nearly elastic SG and ST appears through the fact that the granular temperature in SG is determined by
the shear rate as $T_H\propto \dot{\gamma}^2$, while the temperature in ST is basically  independent of the shear rate.
This conjecture is used in the analysis of the spatial correlations 
in SG \cite{Otsuki09:LR}.

Thus, the solution of eq. (\ref{u_kp})  
with dropping $F(\bv{k},t)$  is given by
\begin{eqnarray}\label{unified-ex}
w^{(i)}_{\bv{k}\perp}(t)&=&w^{(i)}_{\bv{k}\perp}(0)
e^{-\nu^*\sqrt{T_H(0)} B(\bv{k},t)},
\end{eqnarray}
where $B(\bv{k},t) = A_1(t)k^2-A_2(t)k_x 
k_y
+A_3(t)k_x^2$.
Here, $A_1(t)=t(1+\beta t/2)$, $A_2(t)=\dot{\gamma}t^2(1+2\beta t/3)$ and 
$A_3(t)=\dot{\gamma}^2t^3/3(1+3\beta t/4)$ with $\beta=m\nu^*\dot{\gamma}^2/(d\sqrt{T_H(0)})$ for SH,
and
$A_1(t)=t$, $A_2(t)=\dot{\gamma}t^2$ and $A_3(t)=\dot{\gamma}^2t^3/3$ for ST and SG.
The solution of eq. (\ref{P_k}) is similarly obtained as 
$P_{\bv{k}}(t)=P_{\bv{k}(\dot \gamma t)}(0)\exp[-D^*\sqrt{T_H(0)}B(\bv{k},t)]$.

Substituting these solutions into eq. (\ref{C2:eq}) with the aid of 
$T_H\equiv (m/d)\int d\bv{v}_0 (\bv{v}_0-\bv{c})^2f_0(v_0)$, 
$\bv{u}_{\bv{k}}(0)\simeq  (V/N) \bv{v}_0$ and $P_{\bv{k}}(0)\simeq 1$,
we obtain
\begin{equation}\label{result-of-C}
C^{(d)}_\alpha(t)  \propto 
\sqrt{\frac{T_H(t)}{T_H(0)}} \int d\bv{k}
   M_\alpha(\bv{k},t) e^{-B(\bv{k},t)},
\end{equation}
where  we have introduced
\begin{eqnarray}
\fl M_y(\bv{k},t) = 
\frac{k_\perp^2}{k k(-\dot \gamma t)}, \
M_\alpha(\bv{k},t) = 
\frac{k_y^2 k_\alpha^2}{k^2 k_\perp^2} \left ( \frac{k}{k(-\dot \gamma t)} -1 \right ) 
- \dot \gamma t \frac{k_x k_y k_\alpha^2}{k k(-\dot \gamma t) k_\perp^2} + \frac{k^2-k_\alpha^2}{k^2},  
\label{My}
\end{eqnarray}
for $\alpha \neq y$.
Here, we note that the proportional constant in eq. (\ref{result-of-C})
depends on the restitution coefficient $e$ through $\nu^*$,
$D^*$ and $T_H(t)$ in eqs. (\ref{u_kp}) and (\ref{P_k}).
Equations (\ref{result-of-C}) and (\ref{My}) describe VACFs 
for the sheared fluids in all-time region.

In the short-time regime $t < \dot{\gamma}^{-1}$
VACFs, eq. (\ref{result-of-C}),
are reduced to the known
result for the fluid at equilibrium (EQ)
$C^{(d)}_\alpha(t)\sim t^{-d/2}$ for all situations.
On the other hand, 
the long time behaviors of $C_\alpha^{(d)}(t)$ obtained from
eq. (\ref{result-of-C}) 
for $t > \dot{\gamma}^{-1}$, which depend on the situations,
differ from those at EQ.

Let us first consider the asymptotic behaviors of  SH for $t\gg \dot{\gamma}^{-1}$,
where $B(t)$ in eq. (\ref{result-of-C}) is dominated by 
$A_3(t)k_x^2 \simeq \dot{\gamma}^2 \beta t^4 k_x^2/4$.
To keep the contribution of this term we introduce the scaled wave number 
$\bv{k}'$ as
$k_x'\equiv  k_x \sqrt{\beta} \dot{\gamma} t^2$ and $k_{\alpha}'\equiv k_{\alpha}\sqrt{\beta }t$
for $\alpha\ne x$. 
Thus, VACFs  for $t\gg \dot{\gamma}^{-1}$ approximately satisfy
\begin{equation}\label{long-SH}
C^{(2)}_x(t) \sim \dot{\gamma}^{-1} t^{-2},
\qquad
C^{(2)}_y(t)\sim \dot{\gamma}^{-3} t^{-4},
\qquad
C^{(2)}(t)\sim \dot{\gamma}^{-1} t^{-2}/2
\end{equation}
for $d=2$, 
while VACFs for $d \geq 3$ are approximately given by
\begin{equation}\label{long-SH_D3}
C^{(d\geq3)}_\alpha(t) \simeq 
C^{(d\geq3)}(t)\sim \beta^{-d/2+1} \dot{\gamma}^{-1} t^{-d}.
\end{equation}

Next, let us consider the asymptotic behaviors of
ST and nearly elastic SG 
for $t\gg \dot{\gamma}^{-1}$,
where $B(t)$ of eq. (\ref{result-of-C}) is dominated by 
$A_3(t)k_x^2 \simeq \dot{\gamma} t^3 k_x^2/3$.
Similar to the case of SH, we introduce the scaled wave number
$\bv{k}{''}$ as $k_x{''}\equiv k_x \dot\gamma t^{3/2}$ and $k_{\alpha}{''}\equiv k_{\alpha} t^{1/2}$ for $\alpha\ne x$.
Thus, 
two-dimensional VACFs of ST and SG for $t\gg \dot\gamma^{-1}$  are identical to
eq. (\ref{long-SH}), while VACFs  
for $d\ge 3$
are approximately given by
\begin{equation}\label{long-ST&SG_D3}
C^{(d\geq3)}_\alpha(t)\simeq 
C^{(d\geq3)}(t)\sim \dot{\gamma}^{-1} t^{-(d+2)/2}.
\end{equation}

It should be noted that 
the long-time behaviors of $C^{(d)}_y(t)$ for SH, ST, and SG
differ from those of $C^{(d)}_\alpha(t)$  with $\alpha \neq y$
only in the case of $d=2$.
To explain this  different scaling of
$C^{(2)}_y(t)$,
we note that the scaled wave number $k_x'$ or $k_x{''}$
differs from that of $k_\alpha'$ or $k_\alpha{''}$
with $\alpha \neq x$,
and $M_y(\bv{k})$ given in eq. (\ref{My}) is proportional to 
$k_\perp^2 \equiv k^2 - k_y^2$, which reduces to $k_\perp^2 = k_x^2$
for $d=2$, and $k_\perp^2 = k_x^2 + \sum_{\alpha \neq x, y} k_\alpha^{''2}$
for $d \geq 3$.
To demonstrate the difference, let us restrict our interest to ST.
In the vicinity of $k_x=0$, $M_y(\bv{k})$ behaves as $M_y(\bv{k})\sim k_{\perp}^2\sim \{ \sum_{\alpha \neq x, y} k_\alpha^{'2} \}/t$ for $d \geq 3$,
while $M_y(\bv{k})\sim k_{\perp}^2\sim {k_x{''}}^2/ (\dot \gamma t^2)$ for $d=2$,
which lead to the asymptotic behavior of $C_y^{(d\ge 3)}\sim \dot\gamma^{-1}t^{-(d+2)/2}$ and
$C_y^{(2)}(t)\sim \dot\gamma^{-3}t^{-4}$.
Thus, we obtain the characteristic long-time behavior of 
$C^{(2)}_y(t)$ among $C^{(d)}_y(t)$.




To verify the validity of our theoretical predictions 
we perform the simulations of hard spherical particle for $d=2$ and $3$.
In our simulation, 
we first prepare the equilibrium state with the temperature $T_I$
at $t = t_0 < 0$.
For $t>t_0$, we impose the shear flow
by the Lees-Edwards boundary condition to avoid the shear-band instability,
and measure the velocity correlation from $t=0$,
where we have confirmed that the homogeneous shear flow with the temperature $T_s$ is realized.
$m$, $\sigma$, and the temperature $T_s$ at $t=0$ are set to be unity.
Thus, the unit time scale is measured by $\tau_s \equiv \sigma \sqrt{m / T_s}$.
In our simulation, the number of the particles we use
is $65536$ for the calculation of VACFs with
the ensembles of $1600$ different initial conditions for $d=2$. 
The number of the particles we use
is $262144$ for the calculation of VACFs with
the ensembles of $170$ different initial conditions for $d=3$. 
We have already checked that the system size is large enough that
any finite size effects are not observed.
We adopt the area fraction $\nu=\nu_c/2$ with the closest packing fraction of 
particle $\nu_c$ for $d=2$ and $\nu=\nu_c/1.9$ for $d=3$.
The densities $\nu$ we chose might be high, 
but our method can be extended to high density cases
\cite{Otsuki09:LR,OtsukiEPJ}.

\begin{figure}[htbp]
\begin{center}
\includegraphics[height=15em]{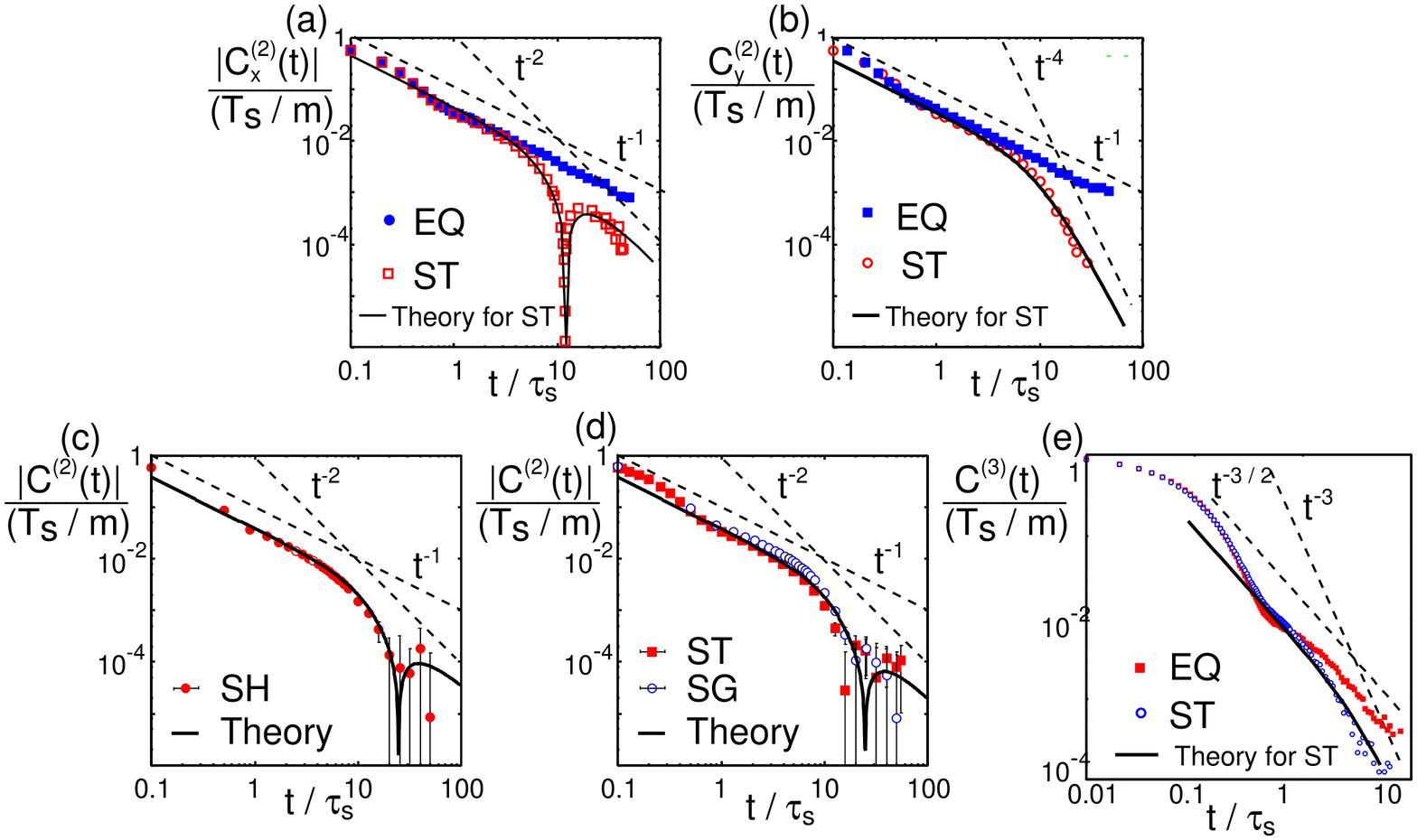}
\caption{
(a) $|C^{(2)}_x(t)|$ as the function of $t$ for ST and EQ with $d=2$.  
(b) $C^{(2)}_y(t)$ as the function of $t$ for ST and EQ with $d=2$.  
(c) $C^{(2)}(t)$ as the function of $t$ for SH with $d=2$.  
(d) $C^{(2)}(t)$ as the function of $t$ for ST and SG
with $d=2$.
(e) $C^{(3)}(t)$ as the function of $t$ for ST and EQ with $d=3$.
Here, we use $N=65636$, $\nu=\nu_c/2$ and $e=0.99$ for SG in (a)--(d), and
$N=262144$, $\nu=\nu_c/1.9$ in (e).
The shear rate is chosen as $\dot{\gamma}=0.2 \tau_s^{-1}$ for SH, ST, and SG
in (a)--(d), and $\dot{\gamma}=0.6 \tau_s^{-1}$ for ST in (e).
}
\label{C_2D}
\end{center}
\end{figure}

We use the parameters  $\dot{\gamma} =0.2 \tau_s^{-1}$ for $d=2$, and $\dot{\gamma}=0.6 \tau_s^{-1}$ 
for $d=3$ in SH, ST, and nearly elastic SG.
In the case of EQ, we use $\dot{\gamma}=0.0$.
For SG, we use the restitution coefficient $e=0.99$
to set the temperature unity.
Here we adopt the velocity rescaling method for ST where the velocity is rescaled for 
every time duration $\Delta t/\tau_s = 0.01$ to keep the constant temperature.


Figures \ref{C_2D} (a)  and (b)  exhibit the numerical results of 
$C_x^{(2)}(t)$ and $C_y^{(2)}(t)$ for EQ and ST with $d=2$.
The theoretical curves in Figs. \ref{C_2D} (a)  and (b) 
  for ST are obtained from eq. (\ref{result-of-C})
with two fitting parameters for the amplitude and the time scale.
$C_x^{(2)}(t)$ and $C_y^{(2)}(t)$ for ST deviate obviously from those for EQ in the long time regime.
These behaviors for ST are almost on the theoretical curves, which strongly supports the validity of
our theory.
We also stress that Fig.  \ref{C_2D} (b) evidently supports the existence of the
theoretical  cross-over of $C_y^{(2)}(t)$ from
$t^{-1}$ to $t^{-4}$, though the numerical cross-over of $C_x^{(2)}(t)$ from $t^{-1}$ to $t^{-2}$ 
in Fig.\ref{C_2D} (a) is obscure. 
From Fig. \ref{C_2D} (a), we confirm that
$|C_x^{(2)}(t)|$ for ST rapidly decreases near $t/\tau_s \simeq 10$
to become negative, while
$C_y^{(2)}(t)$ is always positive, which is consistent with the theoretical
prediction.

Figures \ref{C_2D}(c)--(e)  show the numerical results of 
$C^{(d)}(t)$.
All the theoretical curves are obtained from eq. (\ref{result-of-C})
with 
one fitting parameter for the amplitude.
From Figs. \ref{C_2D} (c) and (d), 
we confirm that VACFs for SH, ST and SG with $d=2$ 
are also consistent with the theoretical result.
From Fig. \ref{C_2D} (e), we find that VACF for ST with $d=3$
deviates from the line $t^{-3/2}$ and VACF for EQ
in the relatively long-time region.
The data are on the theoretical curves again, 
which strongly support the validity of the theory,
although the asymptotic behaviors for $t\gg \dot\gamma^{-1}$  
could not be confirmed.

One may be skeptical about
the condition $\dot\gamma \ll \nu^*\sqrt{T_H(t)} k(-\dot \gamma t)^2$
to ignore $F(\bv{k},t)$ in eq. (\ref{u_kp}), although the agreement between the theory and the simulation is
good.
For more precise analysis,
one can solve the eigenvalue problem of linearized hydrodynamics
around the uniform shear flow \cite{Lutsko,Otsuki09,OtsukiEPJ}.
From the eigenvalue analysis, the transverse mode 
$w^{(i)}_{\bv{k}\perp}(t)$ can be represented by the linear combination of
two eigenvalues $\lambda^{T}(\bv{k}(-\dot \gamma t))$ and 
$\lambda^{S}(\bv{k}(-\dot \gamma t))$
and associated eigenvectors,
where we have introduced
$\lambda^{T}(\bv{k}) =
\nu^* \sqrt{T_H(t)}k^2 - \dot \gamma k_x k_y / k^2$ 
and  
$\lambda^{S}(\bv{k}) =
\nu^* \sqrt{T_H(t)}k^2$.
It should be noted that
the transverse mode
$w^{(1)}_{\bv{k}\perp}(t) = w^{(1)}_{\bv{k}\perp}(0)
\exp[ - \int_0^t ds \lambda^{T}(\bv{k}(-\dot\gamma s)) ]$
always exists, and $w^{(i\ge 2)}_{\bv{k}\perp}(t)$ associated with 
$\lambda^S(\bv{k})$ does not
exist for two dimensional case while $w^{(i\ge 2)}_{\bv{k}\perp}(t)$ is $d-2$ degenerated for $d\ge 3$.
Since $\lambda^S(\bv{k})$ and its associated eigenvectors are identical to those
discussed in eq. (\ref{u_kp}) with neglecting $F(\bv{k},t)$,
we should estimate the contribution from $\lambda^T(\bv{k})$ and $w^{(1)}_{\bv{k}\perp}(t)$.
Substituting $P_{\bv{k}}(t)$ and $w^{(1)}_{\bv{k}\perp}(t)$ with 
$u^{(1)}_{\bv{k}\perp}(t) = \sqrt{T_H(t)} w^{(1)}_{\bv{k}\perp}(t) $
into eq. (\ref{C2:eq}), we obtain its contribution to $C^{(d)}(t)$ as
$C^T(t)\propto
\int d\bv{k} 
\{ (k^2-\dot \gamma t k_x k_y)/k(-\dot \gamma t)^2 \}
\exp[-\int_0^tds\sqrt{T_H(s)}(\nu^*+D^*)k(-\dot\gamma s)^2 ]$,
where we have used $\int_0^tds \dot\gamma k_xk_y(\dot\gamma s)/k(\dot\gamma s)^2=\ln[k(\dot\gamma t)/k]$.
$C^{T}(t)$ obviously satisfies 
$C^{T}(t)\sim t^{-d/2}$ for $t\ll \dot\gamma^{-1}$, 
while the long time
asymptotic form for $t\gg \dot\gamma^{-1}$ is given by
$C^{T}(t)\sim \dot\gamma^{-1} t^{-(d+2)/2}$
for ST, where the proportional constant is 
$\int d\bv{K} [ (K_y^2+K_z^2-K_x K_y) / \{ (K_y-K_x)^2+K_z^2 \} ]
\exp[-\sqrt{T_H(0)}(\nu^*+D^*)(K^2-K_xK_y-2K_x^2/3)]$
for three dimensional case with $K_x=k_x(\dot\gamma t^{3/2})$ and $K_\alpha=k_\alpha t^{1/2}$ for $\alpha\ne x$.
Similarly, the asymptotic forms of $C^{T}(t)$ 
for SH and SG are proportional to those analyzed in the text.
Therefore, we believe that 
the contributions from $F(\bv{k},t)$ is only the change of the amplitude. 
The details of eigenvalue analysis will be discussed elsewhere \cite{OtsukiEPJ}.

Our result is contradicted with Kumaran's prediction \cite{Kumaran,Kumaran09}.
This discrepancy does not disprove Kumaran's scaling,
because ours is for dilute nearly elastic SG but his prediction is only valid for sheared dense granular fluids.
However, we believe that our result is still valid even for dense sheared systems.
Indeed, the universal results between SG and ST
 has been confirmed for the  equal-time long-range  correlation functions \cite{Lutsko,Otsuki09:LR}
even for dense systems, when the systems satisfy the uniform shear. We will discuss the unified description on both the long-range correlation
and the long-tail elsewhere \cite{OtsukiEPJ}.


In conclusion, we have analytically calculated the behaviors of 
VACF in the sheared fluids, which is represented by eq. (\ref{result-of-C}).
(i) We have predicted the existence of the cross-over of VACF from $t^{-d/2}$ to $t^{-d}$ for
sheared heating elastic particles (SH), 
while the cross-over is from $t^{-d/2}$ to $t^{-(d+2)/2}$ for the sheared elastic particles with a thermostat (ST)
and the nearly elastic sheared granular particles (SG) through the hydrodynamic analysis. 
(ii) We have also theoretically predicted that the behavior of VACF for ST 
is almost equivalent to that for SG, while SH is different from that of ST.
(iii) As stated in (ii), we have confirmed that the hydrodynamic properties of dilute and nearly elastic SG are not unusual, which 
is contrast to  Kumaran's prediction  $C(t)\sim t^{-3d/2}$.

%
%
\ack
We thank H. Wada for his useful comments and critical reading of this manuscript.
This work is partially supported by Ministry of Education, Culture, Science and Technology (MEXT), Japan
 (Grant Nos. 21015016, 21540384 and 21540388) and 
 the Grant-in-Aid for the global COE program
"The Next Generation of Physics, Spun from Universality and Emergence"
from MEXT.
The numerical calculations were carried out on Altix3700 BX2 at YITP in Kyoto University.

\end{document}